\documentclass{ifacconf}

\usepackage{graphicx}      % include this line if your document contains figures
\usepackage{verbatim}
\usepackage{amsbsy}
\usepackage[dvips]{epsfig}
\usepackage{epstopdf}
\usepackage{array}
\usepackage{amsmath}
\usepackage{amssymb}
\usepackage{natbib}        % required for bibliography

\begin{document}
\begin{frontmatter}

\title{Model predictive control guided with optimal experimental design for pulse-based parallel cultivation \thanksref{ack}} 
% Title, preferably not more than 10 words.

\author[First]{Jong Woo Kim \thanksref{corr}}
\author[First]{Niels Krausch}
\author[First]{Judit Aizpuru}
\author[Second]{Tilman Barz}
\author[Third]{Sergio Lucia}
\author[Fourth]{Ernesto C. Mart\'inez}
\author[First]{Peter Neubauer}
\author[First,Fifth]{Mariano N. Cruz Bournazou}

\address[First]{Technische Universit{\"a}t Berlin, Chair of Bioprocess Engineering, Strasse des 17. Juni 135, 10623 Berlin, Germany}
\address[Second]{AIT Austrian Institute of Technology GmbH, Giefingasse 2, 1210 Vienna, Austria}
\address[Third]{Technische Universit{\"a}t Dortmund, Department of Biochemical and Chemical Engineering, Emil-Figge-Strasse 70, 44227 Dortmund, Germany}
\address[Fourth]{INGAR (CONICET-UTN), Avellandeda 3657, S3002GJC Santa Fe, Argentina}
\address[Fifth]{DataHow AG, Z{\"u}richstrasse 137, 8600 D{\"u}bendorf, Switzerland}
\thanks[ack]{This work was supported by the German Federal Ministry of Education and Research through the Program International Future Labs for Artificial Intelligence (Grant number 01DD20002A).}
\thanks[corr]{Corresponding author, E-mail: jong.w.kim@tu-berlin.de}

\begin{abstract}                % Abstract of not more than 250 words.
Optimal experimental design for parameter precision attempts to maximize the information content in experimental data for a most effective identification of parametric model. With the recent developments in miniaturization and parallelization of cultivation platforms for high-throughput screening of optimal growth conditions massive amounts of informative data can be generated with few experiments. Increasing the quantity of the data means to increase the number of parameters and experimental design variables which might deteriorate the identifiability and hamper the online computation of optimal inputs. To reduce the problem complexity, in this work, we introduce an auxiliary controller at a lower level that tracks the optimal feeding strategy computed by a high-level optimizer in an online fashion. The hierarchical framework is especially interesting for the operation under constraints. The key aspect of this method are discussed together with an \textit{in silico} study considering parallel glucose limited bacterial fed batch cultivations. 
\end{abstract}

\begin{keyword}
Interaction between design and control, Bio-applications, Batch process modeling and control, Design of experiments
\end{keyword}

\end{frontmatter}
%===============================================================================

\section{Introduction} \label{sec:Intro}
Obtaining a mechanistic model of the microbial system is crucial to the effective, consistent, and reliable bioprocess development \citep{neubauer2013consistent}. A mechanistic model is described by the set of parameters, which characterize the kinetics of the key metabolic pathways \citep{anane2019modelling}. The parameters are estimated by fitting the parameter to the experimental data with the maximum likelihood estimation objective. The accuracy of the mechanistic model is significantly affected by the amount of experimental data available. Nonetheless, it is difficult to obtain the satisfactory amount of data especially for a bioprocess. This is because the cultivation experiment is typically costly and time consuming. Moreover, the key states of the cultivation are mainly measured through the at-line channels, which are significantly scarcer than those from the online sensors.

Recently, high-throughput (HT) technology allows for obtaining massive experimental data, which accelerates bioprocess development \citep{cruz2017online}. Liquid handling station supports for automatizing, parallelizing, and miniaturizing the experimental facilities to perform the laborious cultivation experiments \citep{hemmerich2018microbioreactor}. In the down-scaled cultivation based on the mini-bioreactors, the heterogeneous (or oscillating) operating conditions in the large-scale bioreactors should be simulated \citep{anane2019modelling}. Moreover, the microfluidic device for the continuous feeding on the milliliter scale reactor is technically difficult to achieve \citep{faust2014feeding}. For these reasons, pulse-based glucose feeding has been incorporated for the HT bioprocess development \citep{hans2020automated}. It is crucial to exactly capture the sharp change of states from the pulse-feed by an exact parameter estimation, because the state values right after the pulse-feed are often locate at the boundary of the operating conditions, activating the corresponding inequality constraints.

Although the data amount is amplified thanks to the HT technique and thereby resolve issues rooted from the scarcity of data, it is still crucial to design the experiment that can maximize the information content of the data given the experimental facility. The purpose of the optimal experimental design (OED) is to search for the feeding strategy that maximizes the information content of the measurement, quantified by the Fisher information matrix (FIM) \citep{franceschini2008model}. Through the sequential procedure, which consists of 1) computing new experimental strategy, 2) obtaining data, and 3) estimating the parameters, the parametric uncertainties become progressively reduced \citep{martinez2009design}. The online OED following such procedure has been applied for dynamic systems such as batch fermentation, liquid chromatography, and oxidative reaction \citep{galvanin2009online, telen2014optimal, barz2016real}. 

A critical issue for the online implementation of the OED is the existence of non-identifiable parameters that make the parameter estimation ill-posed. Especially at the beginning of an experiment (i.e., batch phase) the scarcity of at-line measurements of bioprocesses might cause a huge error for the initial parameter guesses, and therefore leads to the failure of the model-based approaches for the entire experiment \citep{lopez2015nonlinear, barz2016real}. Running parallel experiments is advantageous to alleviate this ill-posedness issue in the sense that the amounts of at-line data are multiplied by the number of parallel reactors. Parameter estimation of the parallel experiments has been successfully performed in \citep{anane2019modelling, hans2020automated}. 

Nonetheless, a new problem arises at the OED side from incorporating multiple reactors simultaneously. The online computation or numerical solution of the re-design problem becomes intractable as the number of design variables increases proportionally to the number of parallel bioreactors. Concerning that the typical time interval for online decision of the cultivation is in the order of few minutes, full parallel OED should be relaxed. There exist only a few studies for the parallel OED. One approach is to decompose to the sequential OED for the individual systems by maximizing $k^{\text{th}}$ largest singular value of the FIM of $k^{\text{th}}$ system \citep{galvanin2007model}. This is based on the assumption that the individual optimizer constitutes the global optimizer. Another approach rather focuses on the stabilization of the numerical solution of the parallel OED problem by the subset selection method \citep{cruz2017online, barz2018adaptive}. However, previous studies have not considered the constrained nonlinear programming formulation based on the full discretization of the dynamic systems and sensitivity. Such method can benefit from the efficient gradient-based solver on providing the analytic derivative \citep{bauer2000numerical}. In addition, the parallel OED becomes flexible in incorporating important process constraints as well as being extended to the robust setting or data-driven methods \citep{korkel2004numerical, lucia2014robust, telen2018uncertainty}. 

This paper extends the work of \cite{barz2018adaptive}, in which the adaptive optimal design of 4 parallel fed-batch mini-bioreactors was performed using single shooting optimization. We add an oxygen constraint to the OED which is important for the cell growth by preventing the oxygen limitation condition. To avoid the computational intractability of the constrained OED under the full discretization method, we separate the problem into the unconstrained OED and the additional constrained controller. The auxiliary controller is responsible for tracking the optimal experimental strategy of each bioreactor computed by the unconstrained OED under the oxygen constraint. Auxiliary controllers now concern the dynamics of each single bioreactor, hence they become independent from each other and easily parallelizable. Moreover, the objective function of the tracking problem is convex and the size of the problem is considerably reduced. The parallel OED is solved with much lower frequency, while the auxiliary controllers are solved in every decision time steps. This hierarchical structure enables the online implementation of the parallel constrained OED by experiencing minimal approximation, as demonstrated similarly in the adaptive optimization of bioprocess \citep{kim2021model}. Proposed methods are demonstrated using the 8 parallel cultivation experiments base on the \textit{in silico} setting. 

\section{Macro-kinetic growth model}
\label{sec:MBR}
The robotic facility of the HT platform is able to conduct eight parallel cultivations. The following measurements are considered:
\begin{itemize}
    \item Online measurements: Dissolved oxygen tension (DOT) and pH are recorded every 30 seconds online. 
    \item Atline measurements: Biomass, substrate, acetate, and product concentrations are analyzed from  samples taken every 120 minutes.
    \item Pipette actions: Glucose solution 200 $g/L$ is added every 10 min in each reactor. The amount of pulse addition of the glucose solution is the decision variable.
\end{itemize}
The cultivation is divided into two phases; batch and fed-batch. In the batch phase, the biomass grows by consuming the substrate which initially exists in the reactor. Fed-batch starts as soon as the substrate is depleted, which can be detected by observing the steep increase in the DOT signal. 

The macro-kinetic growth (MKG) model is described by a set of ordinary differential equations (ODEs) of six state variables, biomass $X$, substrate $S$, acetate $A$, dissolved oxygen tension measurement $DOT_m$, product $P$, and the reactor medium volume $V$. The governing equations for the biomass, substrate, acetate, and product concentrations are expressed as
\begin{align} \label{eq:MKG_concentrations}
    \dfrac{dX}{dt} &= -\mu X + \dfrac{X}{V}  F_{\lambda}\\
    \dfrac{dS}{dt} &= -q_S X + \dfrac{S}{V}  F_{\lambda}\\ 
    \dfrac{dA}{dt} &= q_A X + \dfrac{A}{V}  F_{\lambda}\\
    \dfrac{dP}{dt} &= q_P X + \dfrac{P}{V}  F_{\lambda}
\end{align}
where $\mu$, $q_S$, $q_A$, and $q_P$ are the specific growth rate ($g/(g\cdot h)$), specific substrate uptake rate ($g/(g\cdot h)$), specific acetate production rate ($g/(g\cdot h)$), and specific product formation rate ($g/(g\cdot h)$), respectively; $F_{\lambda}$ ($L/h$) is the evaporation rate. Dissolved oxygen tension (DOT) is modelled by the algebraic equation considering equilibrium oxygen concentration in the reactor medium as
\begin{equation} \label{eq:MKG_DOTalg}
DOT = DOT^* - \dfrac{q_OXH}{k_{la}}
\end{equation}
where $k_{la}$ ($h^{-1}$) denotes the volumetric oxygen transfer coefficient; $DOT^*$ ($\%$) denotes the saturation concentration of DOT; $q_O$ ($g/(g\cdot h)$) denotes the specific oxygen uptake rate; $H$ ($mol/(m^3 \cdot Pa)$) denotes the Henry constant. The oxygen sensor has first order delay, which is written as
\begin{equation} \label{eq:MKG_DOTm}
\dfrac{dDOT_m}{dt} = k_{p}(DOT - DOT_m)
\end{equation}
where $k_p$ ($h^{-1}$) represent the time constant.

The parameter vector $\theta$ comprises of the physical parameters of the MKG model. The reader is referred to \cite{anane2017modelling} for the detailed description of the MKG model based on the acetate cycling and glucose partitioning. We distinguish the global parameters and local reactor-dependent parameters as $\theta_g$ and $\theta_{l}$, respectively as;
\begin{equation} \label{eq:MKG_parameters}
\begin{split}
    \theta_g &= \left\lbrace \begin{array}{l}
            q_{S, max}, q_m, q_{Ap, max}, q_{Ac, max},  \\
           Y_{XS, em}, Y_{AS, of}, Y_{XA}, Y_{OS}, Y_{OA}, Y_{PS}, \\
           K_S, K_{qS}, K_{i, SA}, K_A, K_{i, AS}, d_{S, ox, P}
    \end{array}  \right\rbrace \\
    \theta_{l} &= \left\lbrace  k_{la}, k_{p} \right\rbrace 
\end{split}
\end{equation}

In the fed-batch phase, the glucose feed is given in pulse-type rather than continuous-type. This results in instantaneous jumps in the process variables, which are governed by the mass balance. Denote $t$ and $t^+$ the times just before and after the pulse input occurs, respectively. The mass balance is described by
\begin{subequations} \label{eq:MKG_MB}
\begin{align}
X(t^+) &= X(t) - \dfrac{X(t)}{V(t^+)}\Delta v(t) \\
S(t^+) &= S(t) - \dfrac{S(t) - S_f}{V(t^+)}\Delta v(t) \\ 
A(t^+) &= A(t) - \dfrac{A(t)}{V(t^+)}\Delta v(t) \\
DOT_m(t^+) &= DOT_m(t) \\
P(t^+) &= P(t) - \dfrac{P(t)}{V(t^+)}\Delta v(t) \\
V(t^+) &= V(t) + \Delta v(t)
\end{align}
\end{subequations}
where $\Delta v(t)$ ($L$) denotes the amount of pulse-feed at time $t$; $S_f$ ($g/L$) denotes the substrate concentration in the pulse-feed.

\section{Model based optimal experimental design}
\subsection{Problem description and formulation}
The states $x$, manipulated variables $u$, and measured variables $y$ comprise of 
\begin{equation}
\begin{split}
x &= \left[ X, S, A, DOT_m, P, V \right] \\
y &= \left[ X, S, A, DOT_m, P \right] \\
u &= \left[ \Delta v \right]
\end{split}
\end{equation}

The high-throughput experiment is characterized by following discrete variables:
\begin{equation}
\begin{split}
    \mathcal{R}&=\left\lbrace (row) | row \in \left\lbrace A, B, \ldots, H \right\rbrace \right\rbrace \\
    \mathcal{U}&=\left\lbrace 10k \text{ (min)}| k\in \mathbb{N} \right\rbrace \\
    \mathcal{M}_{r, y} &=\left\lbrace \begin{array}{l} 
    \begin{split}
        \left\lbrace 120k \text{ (min)} | k \in \mathbb{N} \right\rbrace, \quad y\in \left\lbrace X, S, A, P \right\rbrace
    \end{split}
         \\
        \left\lbrace 30k \text{ (sec)} | k \in \mathbb{N} \right\rbrace, \quad  r\in \mathcal{R}, \quad y= DOT_m 
        \end{array} \right. 
\end{split}
\end{equation}
where $\mathcal{R}$ is the index set of the mini-bioreactors; $\mathcal{U}$ is the discrete pulse-feeding times; $\mathcal{M}_{r, y}$ is the measurement times of the reactor $r$ for the measured variable $y$. The collection of all-time elements of $\mathcal{M}_{r, y}$ is denoted as $\mathcal{M}=\bigcup_{y, r\in \mathcal{R}}\mathcal{M}_{r, y}$.

Denote differential equations of the MKG model (Eqs.~(\ref{eq:MKG_concentrations})-(\ref{eq:MKG_DOTm})) as $f \in \mathbb{R}^{n_x}$, algebraic equations of mass balance due to the pulse-feed (Eqs.~(\ref{eq:MKG_MB})) as $f_d \in \mathbb{R}^{n_x}$, and output functions for reactor $r$ as $h_r \in \mathbb{R}^{n_y}$. Then the parallel cultivation setup of the mini-bioreactors can be described in the compact form as:
\begin{equation} \label{eq:MKGMBRCompact}
\begin{split}
\dot{x}_r(t) &= f(x_r(t), \theta_r), \quad t \in [t_0, t_f] \setminus \mathcal{U} \\
x_r(t^+) &= f_d(x_r(t), u_r(t)), \quad t\in \mathcal{U} \\
y_r(t) &= h_r(x_r(t)) \quad t \in \mathcal{M} \\
x_r(t_0) &= x_{0, r}, \quad \forall r \in \mathcal{R}
\end{split}
\end{equation}
where the subscript $r$ indicates that variables $x$, $u$, and $y$ belong to the reactor $r$; $t_0$ and $t_f$ are the initial and final cultivation time, respectively; $t^+$ is the time after which pulse-feed is made. The parameter vector for the individual reactor $r\in \mathcal{R}$ is denoted as $\theta_r$, and the parameter vector that contains the global parameters and the local parameters for the entire reactors is denoted as $\theta \in \mathbb{R}^{n_{\theta}}$. Following shows the definition:
\begin{equation}
\begin{split}
\theta_r &= \left[\theta_g, \theta_{l, r} \right], \quad r \in \mathcal{R} \\
\theta &= \left[\theta_g, \left\lbrace \theta_{l, r} | r \in \mathcal{R} \right\rbrace \right]
\end{split}
\end{equation}
Each bioreactor has the initial condition $x_{0, r}$. The output function is given by $h_r(x_r(t)) = x_r(t)$ if $t \in \mathcal{M}_{r, y}$ and not defined elsewhere.

\subsection{Parameter estimation and Fisher information matrix}
\label{sec:MHE}

Parameters are estimated based on the maximum likelihood estimation method by incorporating previous measurements. Measurements at time instance $t_i$ are assumed to follow the normal distribution with the variance-covariance matrix $\Sigma_i \in \mathbb{R}^{n_y \times n_y}$. The correlations between measurements are not considered, hence $\Sigma_i$ is a diagonal matrix. The objective function for the reactor $r$ is given as follows:
\begin{equation} \label{eq:MHEErrorPerReactor}
\begin{split}
E_r(\theta) &= \dfrac{R_1}{2}\sum_{t_i \in \mathcal{M}} (y_r(t_i, \theta) - y_{r, i})^T \Sigma_i^{-1} (y_r(t_i, \theta) - y_{r, i})
\end{split}
\end{equation}
Here $R_1$ is a weighting constant; $y_r(t_i, \theta)$ is the output at time $t_i$ predicted from the model with parameter $\theta$; $y_{r, i}$ is the respective measurement. The residual vector is scaled by $|\mathcal{M}_{r, y}|$, the number of measurements on the reactor $r$ and output variable $y$. Parameters are estimated from the total objective function defined as:
\begin{equation} \label{eq:MHEobjective}
\begin{split}
E(\theta) &= \min_{\theta} \| \hat{\theta}_{g, \text{prev}} - \theta_g \|_{R_0}^2 \\
&+ \sum_{r\in \mathcal{R}} \left( \| \hat{\theta}_{l, r, \text{prev}} - \theta_{l, r} \|_{R_0}^2 +  E_r(\theta) \right)
\end{split}
\end{equation}
where $R_0$ denotes a weighting constant. Eq.~(\ref{eq:MHEobjective}) is the summation of Eq.~(\ref{eq:MHEErrorPerReactor}) with respect to all reactors. Here, Tikhonov regularization that penalizes the parameters deviating from the previous estimates (i.e., $\hat{\theta}_{g, \text{prev}}$ and $\hat{\theta}_{l, r, \text{prev}}$) is implemented to prevent the ill-conditioning issue \citep{barz2016real}. The optimal parameter vector $\hat{\theta}$ is evaluated by minimizing $E(\theta)$ subject to the model equations in Eqs.~(\ref{eq:MKGMBRCompact}).

If $\hat{\theta}$ is the unconstrained optimizer of Eq.~(\ref{eq:MHEobjective}), then the covariance matrix can be approximated according to the Cramer-Rao inequality as
\begin{equation}
\left( C_{[0, t_0]}(\hat{\theta}) \right)^{-1} \approx \sum_{r\in \mathcal{R}}\sum_{t_i\in \mathcal{M}} s_r(t_i)^T \Sigma_i^{-1} s_r(t_i)
\end{equation}
where $s_r(t_i) \in \mathbb{R}^{|\mathcal{M}| \times n_{\theta}}$ is the sensitivity matrix of the output of the bioreactor $r$ with respect to the parameter vector, given as
\begin{equation}
s_r(t_i) = \dfrac{\partial y_r}{\partial \theta}
\end{equation}
where $\left( C_{[0, t_0]}(\hat{\theta}) \right)^{-1}$ is referred to as Fisher information matrix, which provides the upper bound of the parameter covariance matrix with the measurement data from time 0 to $t_0$. 

% Identifiability analysis is performed with the sensitivity matrix $s_r$. The singular values of the column stack of the sensitivity matrices for all reactors,  $S_{\mathcal{R}}=[s_r]_{r\in \mathcal{R}}\in \mathbb{R}^{|\mathcal{R}||\mathcal{M}| \times n_{\theta}}$, is the common criteria for selecting the subset of parameters to be estimated. We use the numerical $\epsilon$-rank for automatically distinguishing the identifiable parameters and their number \citep{lopez2015nonlinear}, computed by
% \begin{equation}
% \epsilon = \max \left[ \dfrac{\zeta_1}{\kappa_{max}(S_{\mathcal{R}})}, 
% \dfrac{1}{\gamma_{\max}(S_{\mathcal{R}})}\right]
% \end{equation}
% where $\zeta_i$ is the $i^{\text{th}}$ largest singular value of $S_{\mathcal{R}}$; $\kappa_{i}=\zeta_1/\zeta_i$ is the sub-condition number; $\gamma_i=1/\zeta_i$ is the sub-collinearity index. The rank of the parameters are determined based on the QR decomposition of $S_{\mathcal{R}}$ \citep{barz2018adaptive}. 

\subsection{Optimal experimental design}
We formulate the OED with the Fisher information matrices (FIM) for the past and future measurements. Denote the current and future times as $t_0$ and $t_p$, respectively, and the collection of the measurement times within $[t_0, t_p]$ as $\mathcal{M}^+$. Under the last estimates $\hat{\theta}$, the Fisher information matrix is additive:
\begin{equation} \label{eq:FIM_future}
\begin{split}
\left( C_{[0, t_p]}(\hat{\theta}) \right)^{-1} &\approx \left( C_{[0, t_0]}(\hat{\theta})  \right)^{-1} \\
&+ \sum_{r\in \mathcal{R}}\sum_{t_i\in \mathcal{M^+}} s_r(t_i)^T \Sigma_i^{-1} s_r(t_i)
\end{split}
\end{equation}
The exact evaluation of Eq.~(\ref{eq:FIM_future}) is impossible, due to the dependency of the FIM on the model parameter values with the measurements, which cannot be known in advance. Therefore, OED approximates the parameters to be fixed to their prior estimates until the prediction horizon $t_p$ \citep{galvanin2009online}.

There exist several scalar metrics for the Fisher information matrix. In this paper, we apply $A$ criterion of the objective function of the OED problem. The definition is given as:
\begin{subequations} \label{eq:OEDcriteria}
\begin{align}
\psi_A &= \dfrac{1}{n_{\theta}} \text{Tr}\left(  C_{[0, t_p]}(\hat{\theta}) \right) 
% \psi_E &= -\lambda_{\text{min}}\left( \left( C_{[0, t_p]}(\hat{\theta}) \right)^{-1} \right)
\end{align}
\end{subequations}
where $\text{Tr}(\cdot)$ stands for the trace of the matrix, respectively.

Dynamic propagation of the sensitivity matrix can be computed from the chain rule as
\begin{equation}
\dot{s}_r(t) = \dfrac{\partial f}{\partial x} s_r(t) + \dfrac{\partial f}{\partial \theta}, \quad s_r(0) = \dfrac{\partial x_r(0)}{\partial \theta}
\end{equation}
In addition, the relationship between sensitivity matrices before and after the pulse-feed is obtained from differentiating the mass balance equations $f_d$ as
\begin{equation}
s_r(t^+) = \dfrac{\partial f_d}{\partial x} s_r(t) + \dfrac{\partial f_d}{\partial \theta}
\end{equation}

Accommodating the dynamic models and dynamic sensitivity matrices for all $r\in \mathcal{R}$, and the objective function, the OED problem optimizes the manipulated variable $u_r, r\in \mathcal{R}$ in $[t_0, t_p]$. The problem is formulated as
\begin{subequations} \label{eq:OEDproblem}
\begin{align}
\min_{u_r, r\in \mathcal{R}} &\quad \psi_A \\
\text{s.t.}  
& \quad \dot{x}_r(t) = f(x_r(t), \theta_r), \quad t \in [t_0, t_p] \setminus \mathcal{U} \label{eq:CultOptProb_MKG}\\
& \quad x_r(t^+) = f_d\left(x_r(t), u_r(t)\right), \quad t\in \mathcal{U} \label{eq:CultOptProb_MB}\\
& \quad \dot{s}_r(t) = \dfrac{\partial f}{\partial x} s_r(t) + \dfrac{\partial f}{\partial \theta}, \quad t \in [t_0, t_p] \setminus \mathcal{U} \\
& \quad s_r(t^+) = \dfrac{\partial f_d}{\partial x} s_r(t) + \dfrac{\partial f_d}{\partial \theta}, \quad t\in \mathcal{U} \\
& \quad x_{\text{min}} \leq x_r(t) \leq x_{\text{max}}, \quad t \in [t_0, t_p]\\
& \quad u_{\text{min}} \leq u_r(t) \leq u_{\text{max}}, \quad t\in \mathcal{U}\\
& \quad x_r(0)=x_{0, r} \\
& \quad r\in \mathcal{R} \label{eq:CultOptProb_MBR}
\end{align}
\end{subequations}
where $x_{\text{min}}$ and $x_{\text{max}}$ are the lower and upper bounds for the state variables; $u_{\text{min}}$ and $u_{\text{max}}$ are the lower and upper bounds for the manipulated variable. 

% The OED with the $E$-criterion objective can be reformulated by introducing an auxiliary parameter $\tau$ to the objective, and adding a matrix inequality constraint, $\left( C_{[0, t_p]}(\hat{\theta}) \right)^{-1} - \tau I \succeq 0$. This can be further modified by means of the Sylvester criterion; a matrix is positive definite if the determinant of all its principal minors are positive \citep{telen2015differentiable}. As a result, following $n_{\theta}$ constraints are added to the original problem Eqs.~(\ref{eq:OEDproblem}) as:
% \begin{subequations}
% \begin{align}
% \max_{u_r, r\in \mathcal{R}} &\quad \tau \\
% \text{s.t.}  
% & \quad \text{Eqs}.~(\ref{eq:CultOptProb_MKG}) - (\ref{eq:CultOptProb_MBR}) \\
% & \quad \begin{array}{l}
%      \det \left( (C_{[0, t_p]}(\hat{\theta})^{-1}  -\tau I )[(1:i) \times (1:i)]\right)>0, \\ \quad i \in \left\lbrace 1, \ldots, n_{\theta} \right\rbrace  
% \end{array} \end{align}
% \end{subequations}

\subsection{OED-guided auxiliary model predictive controller}
The OED problem in Eqs.~(\ref{eq:OEDproblem}) comprises the model and sensitivity dynamics from all bioreactors. Thus it is intractable to perform the online re-design of 8 bioreactors. However, if the experimental design is not adapted to recent measurements, the process will be operated under the region with low information content. This cannot prevent the ill-conditioning problem of the parameter estimation, and eventually leads to the violation of the crucial DOT constraint, which is parameter dependent (see Eq.~(\ref{eq:MKG_DOTalg})).

\begin{figure}
\centering
\includegraphics[width=\linewidth]{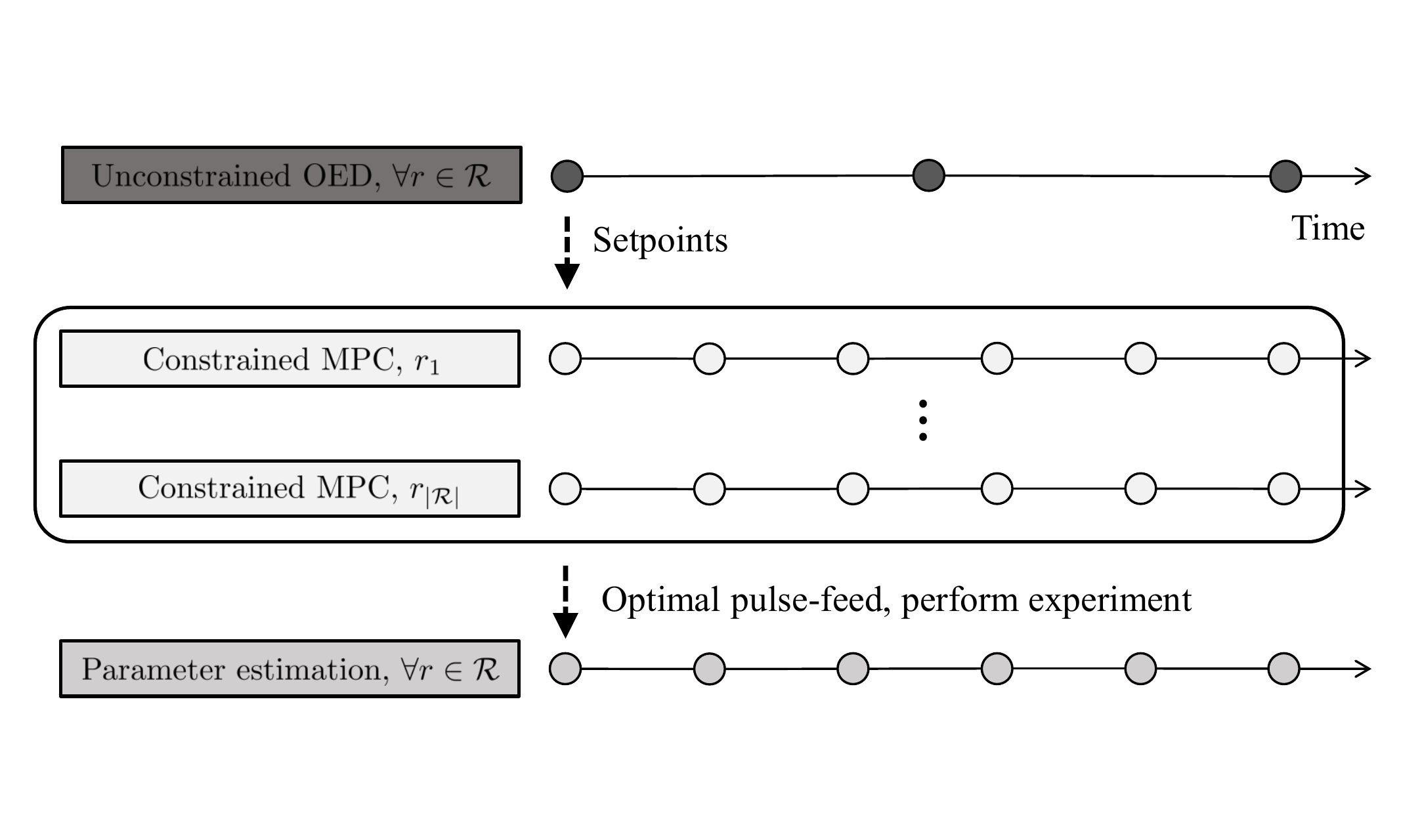} 
\caption{Hierarchical structure of the parallel OED and auxiliary MPC controllers.}
\label{fig:OED_MPC_scheme}
\end{figure}

To overcome such limitation, we propose utilizing an auxiliary model predictive controller (MPC) that tracks the pulse-feed strategy of a single bioreactor computed by the OED and considering the constraints in the auxiliary MPC instead of in the parallel OED. The information content increases roughly along the amount of the feed, because magnitudes of the state jumps are proportional to the feed amount (see Eqs.~(\ref{eq:MKG_MB})). The oxygen uptake becomes larger at the same time, hence the solution lies at the boundary of the constraint. In this case, it is difficult to obtain the converged solution of the optimization problem with the nonconvex the objective function. Therefore, we relax the constraint of the OED problem and move it to auxiliary controllers. The optimization problem for the auxiliary controller for the reactor $r\in \mathcal{R}$ is formulated as follows:
\begin{subequations} \label{eq:CultOptProb}
\begin{align}
\min_{u_r} &\quad \sum_{t\in \mathcal{U}} \| u_r(t) - u_r^{oed}(t) \|^2  \\
\text{s.t.}  
& \quad \dot{x}_r(t) = f(x_r(t), \theta_r), \quad t \in [t_0, t_p] \setminus \mathcal{U} \\
& \quad x_r(t^+) = f_d\left(x_r(t), u_r(t)\right), \quad t\in \mathcal{U} \\
& \quad x_{\text{min}} \leq x_r(t) \leq x_{\text{max}}, \quad t \in [t_0, t_p] \setminus \mathcal{U} \\
& \quad u_{\text{min}} \leq u_r(t) \leq u_{\text{max}}, \quad t\in \mathcal{U}\\
& \quad DOT_{lb, r} \leq DOT_r(t, \theta_r), \quad t \in [t_0, t_p] \\
& \quad x_r(0)=x_{0, r},
\end{align}
\end{subequations}
where $u_r^{oed}(t)$ is the pulse-feed strategy computed from the optimization problem of Eqs.~(\ref{eq:OEDproblem}); $DOT_{lb, r}$ represents the DOT lower bound of the reactor $r$. In the auxiliary controller, not only a simple quadratic function is used, but also the dynamic sensitivity equations with large number of decision variables no longer exist. Moreover, the parallelization is straightforward because only a single bioreactor is considered per each controller. The implementation procedure is illustrated in Fig.~\ref{fig:OED_MPC_scheme}. Unconstrained OED is computed with lower frequency due to its complexity, while auxiliary MPC controllers for the individual reactors are computed every decision time steps (i.e., 10 min). With this hierarchical structure, the experimental design for the parallel mini-bioreactors with the oxygen constraint can be conducted online.

\section{\textit{In silico} results}

The \textit{in silico} cultivation for 8 mini-bioreactors is performed. To generate the \textit{in silico} data, we add uncertainties to the parameters by adding random uniform noise with 10 \% support to their scale. Moreover, Gaussian noise with 5 \% variance to the scales is added to each measurement. Do-mpc software, a toolbox developed based on CasADi \citep{Andersson2019}, is utilized to solve the optimization problems \citep{lucia2017rapid}. The prediction horizon for the OED and the auxiliary MPCs are until the end of the batch and 180 min, respectively. 

\begin{figure}
\centering
\includegraphics[width=\linewidth]{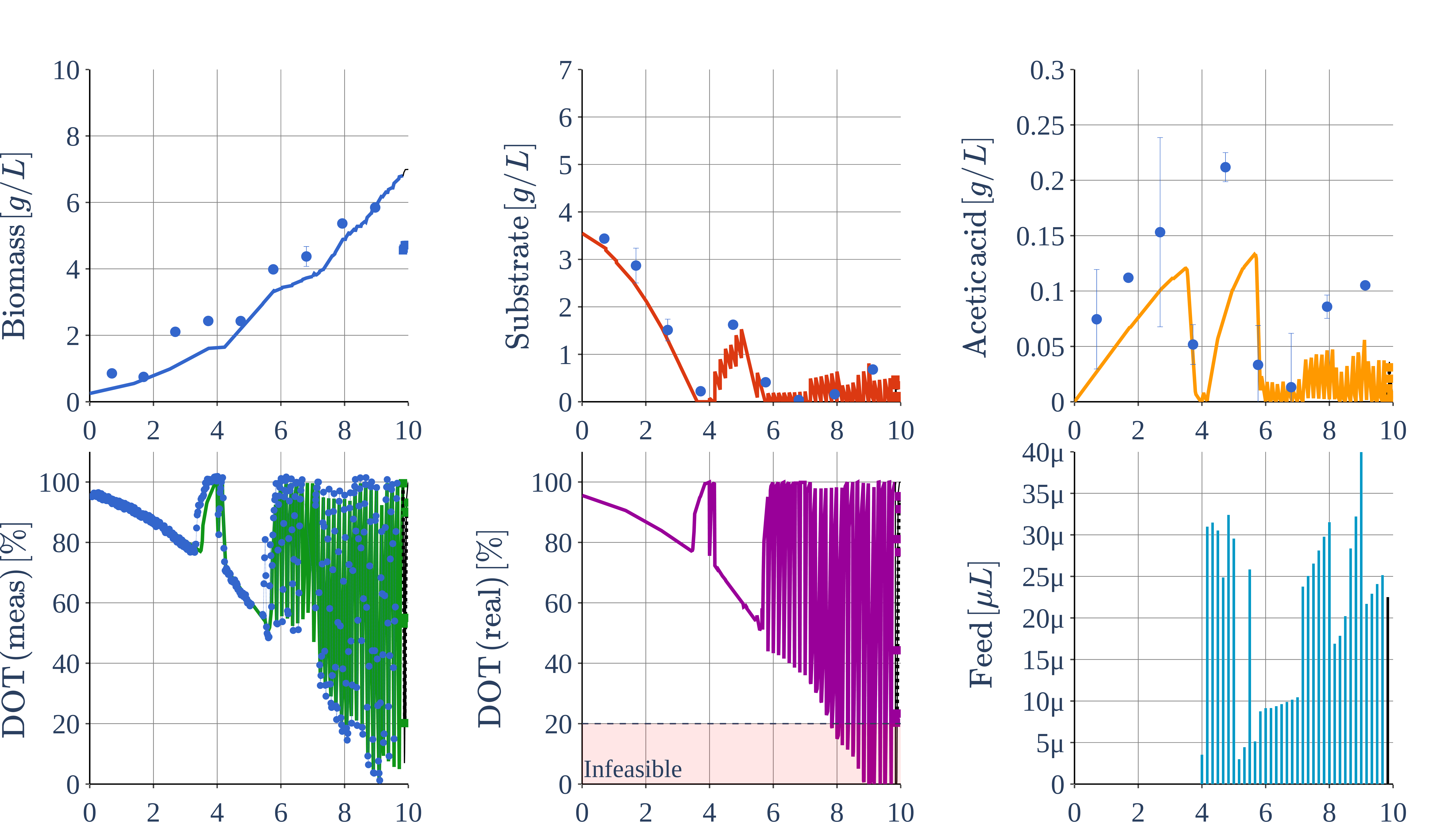} 
\caption{State trajectory under the pulse-feed computed by the pure OED. The colored solid line represents the state trajectory, the dot represents the measurement.}
\label{fig:OEDresult}
\end{figure}

\begin{figure}
\centering
\includegraphics[width=\linewidth]{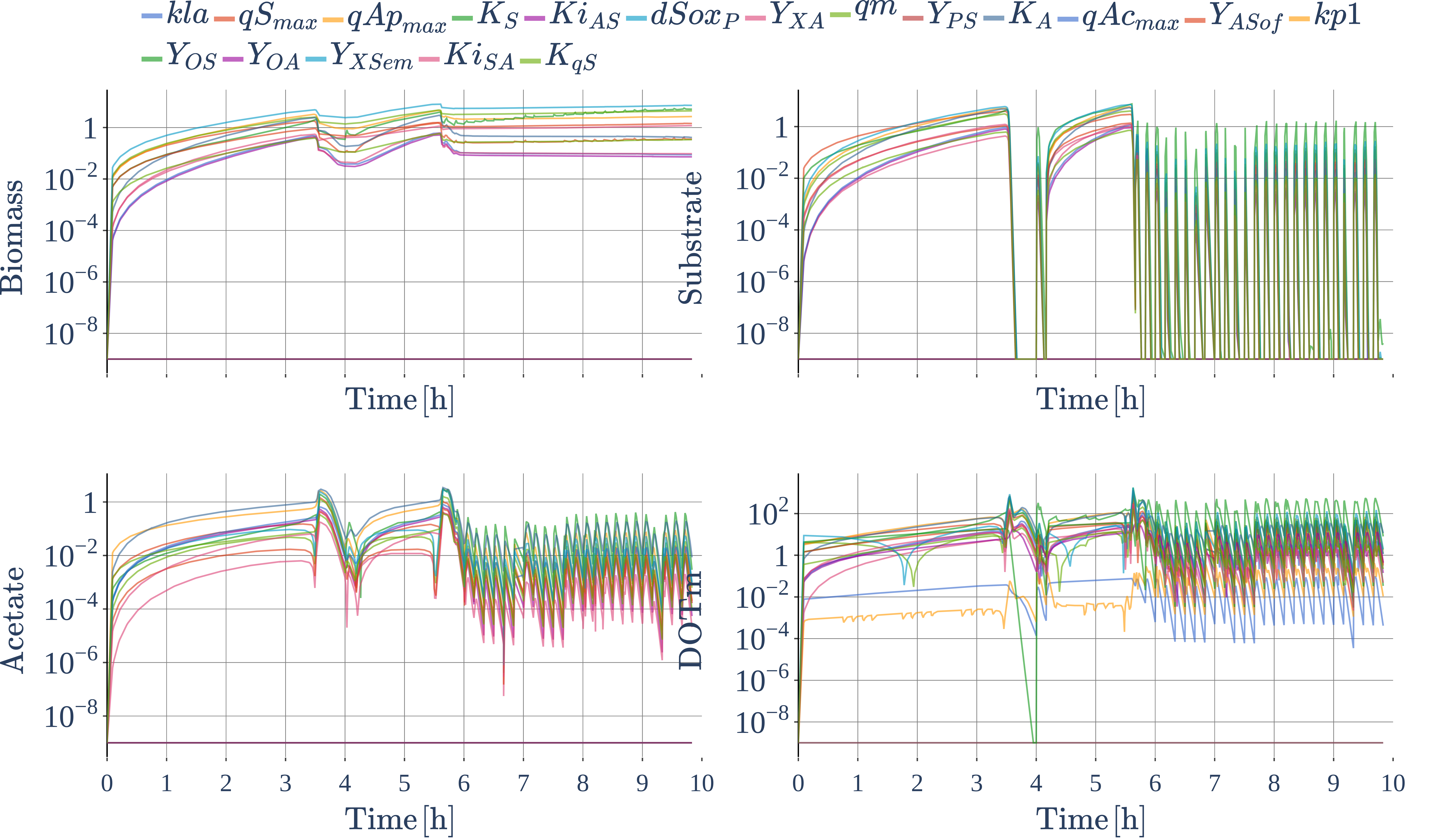} 
\caption{Dynamic sensitivities for four states with respect to 18 parameters throughout the cultivation experiment.}
\label{fig:Sensitivity}
\end{figure}

Cultivation results of one of eight reactors operated by the unconstrained OED without auxiliary controllers is presented in Fig.~\ref{fig:OEDresult}. Biomass, substrate, acetate, $DOT_m$, $DOT$, and pulse-feed trajectories are plotted. The OED problem is solved every 60 min, due to its high computational cost. The sensitivities for the four states (i.e., $X$, $S$, $A$, and $DOT_m$) with respect to 18 kinetic parameters $\theta_g$ and $\theta_1$ (Eq.~(\ref{eq:MKG_parameters})) are shown in Fig.~\ref{fig:Sensitivity}. This figure highlights that the magnitude of sensitivity values arise as the pulse-feed is given, similar to the dynamic behavior of the state values from the mass balance equations. The increase of sensitivities are proportional to the feed amount, and therefore the OED objective encourages to feed as much as possible. Since the oxygen uptake is proportional to the biomass amount according to the Eq.~(\ref{eq:MKG_DOTalg}), the optimal solution of the OED problem activates the DOT constraint.

\begin{figure}
\centering
\includegraphics[width=\linewidth]{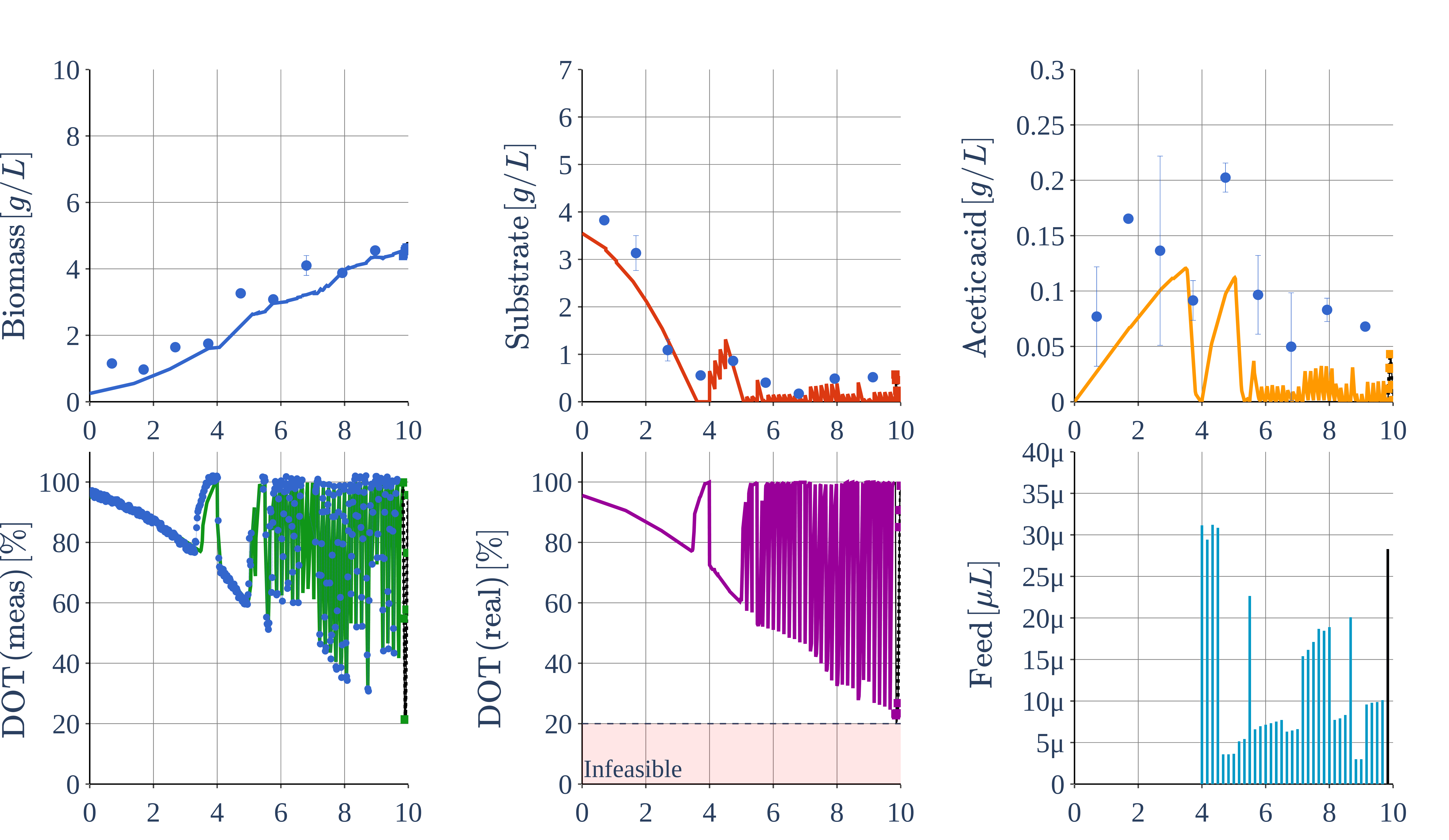} 
\caption{State trajectory under the pulse-feed computed by OED guided MPC. The colored solid line represents the state trajectory, the dot represents the measurement.}
\label{fig:OED_MPCresult}
\end{figure}

Figure~\ref{fig:OED_MPCresult} depicts the cultivation results of one of eight reactors operated by the OED problem with the auxiliary MPC. The MPC is solved every 10 min, and it is parallelized to the problem for each reactor. MPC tracks the OED result initially where the violation does not happen. After the violation is detected within the prediction horizon, the pulse-feed amount is decreased. The feasibility can be guaranteed by shifting the DOT constraint to the auxiliary controllers. 

Finally, we validate that the loss of the information content of the pulse-feed strategy computed by the proposed method is not significant compared to the constrained OED. For comparison, we consider the OED problem for a single reactor because the optimization problem is able to be solved online. $\psi_A$ values for three different feeding strategies, proposed hierarchical structure, constrained OED problem, and the OED without constraint, are 1.014, 1.106, and 3.386, respectively. This indicates that due to the approximation the reduction of the information content is unavoidable, however not considerable. 

\section{Concluding remarks}
In this study, we introduce the auxiliary controller to enable online implementation of the computationally intractable OED for the parallel cultivation. The auxiliary controller has a simple structure that can be parallelizable, and has a quadratic objective for tracking the results from the OED problem. The additional constraint such as oxygen limitation constraint can be relaxed from the OED problem and then be accounted for to the auxiliary controller. Through the \textit{in silico} study for the cultivation of eight parallel replicates, we demonstrate that the proposed approach yields a closed-loop feeding strategy with near-optimal information content within the feasible region. Future work will focus on extending the strategy to the full high-throughput experiment involving different experimental conditions. Results obtained demonstrate that it is also important to account for the parametric uncertainty by formulating robust OED.

\bibliography{ifacconf}             % bib file to produce the bibliography
                                                     % with bibtex (preferred)
                                                   
%\begin{thebibliography}{xx}  % you can also add the bibliography by hand

%\bibitem[Able(1956)]{Abl:56}
%B.C. Able.
%\newblock Nucleic acid content of microscope.
%\newblock \emph{Nature}, 135:\penalty0 7--9, 1956.

%\bibitem[Able et~al.(1954)Able, Tagg, and Rush]{AbTaRu:54}
%B.C. Able, R.A. Tagg, and M.~Rush.
%\newblock Enzyme-catalyzed cellular transanimations.
%\newblock In A.F. Round, editor, \emph{Advances in Enzymology}, volume~2, pages
%  125--247. Academic Press, New York, 3rd edition, 1954.

%\bibitem[Keohane(1958)]{Keo:58}
%R.~Keohane.
%\newblock \emph{Power and Interdependence: World Politics in Transitions}.
%\newblock Little, Brown \& Co., Boston, 1958.

%\bibitem[Powers(1985)]{Pow:85}
%T.~Powers.
%\newblock Is there a way out?
%\newblock \emph{Harpers}, pages 35--47, June 1985.

%\bibitem[Soukhanov(1992)]{Heritage:92}
%A.~H. Soukhanov, editor.
%\newblock \emph{{The American Heritage. Dictionary of the American Language}}.
%\newblock Houghton Mifflin Company, 1992.

%\end{thebibliography}

\end{document}